Creativity

Liane Gabora

Note: This draft may not be identical to the version that was submitted to Oxford University Press.

**Table of Contents**

**Summary**

Creativity is perhaps what most differentiates humans from other species. Understanding creativity is particularly important in times of accelerated cultural and environmental change such as the present, in which novel approaches and perspectives are needed. The study of creativity is an exciting area that brings together many different branches of psychology: cognitive, social, personality, developmental, organizational, clinical, neuroscience, mathematical models, and computer simulations. The creative process is thought to involve the capacity to shift between divergent and convergent modes of thought in response to task demands. Divergent thought is conventionally characterized as and the kind of thinking needed for open-ended tasks, and measured by the ability to generate multiple solutions, while convergent thought is commonly characterized as the kind of thinking needed for tasks in which there is only one correct solution. More recently, divergent thought has been conceived of as reflecting on the task from unconventional contexts or perspectives, while convergent thought has been conceived of as reflecting on it from conventional contexts or perspectives. Personality traits correlated with creativity include openness to experience, tolerance of ambiguity, impulsivity, and self-confidence. Evidence that creativity is linked with affective disorders is mixed. Neuroscientific research on creativity using electroencephalography (EEG) or functional magnetic resonance imaging (fMRI) suggest that creativity is associated with a loosening of cognitive control and decreased arousal. It has been shown that the distributed, content-addressable structure of associative memory is conducive to bringing task-relevant items to mind without the need for explicit search. Tangible evidence of human creativity date back to the earliest stone tools over three million years ago, with the Middle-Upper Paleolithic marking the onset of art, science and religion, and another surge of creativity in the present. Past and current areas of controversy concern the relative contributions of expertise, chance, and intuition, whether the emphasis should be on process versus product, whether creativity is domain-specific versus domain-general, the extent to which creativity is correlated with affective disorders, and whether divergent thinking entails the generation of multiple ideas or the honing of a single initially ambiguous mental representation that may manifest as different external outputs. Promising areas for further psychological study of creativity include computational modeling, research on the biological basis of creativity, and studies that track specific creative ideation processes over time.

**What *is* Creativity?**

Creativity is arguably our most uniquely human trait, yet it is perhaps the most elusive aspect of the human condition. When we create, we may feel deeply alive in the present moment, yet in another sense feel as if we are living *beyond* the present moment, oblivious to what is happening around us as we weave together new understandings of the past or new possibilities for the future.

Defining creativity is difficult; for example, not all creative outputs are aesthetically pleasing, and not all are useful, though both aesthetic value and usefulness capture, in some sense, what creativity is about. There are cultural differences in the conceptualization of creativity (Leung, 2018). Nevertheless, psychologists have converged on the definition originally proposed by Guilford (1950) over half a century years ago. Guilford defined creativity in terms of two criteria: *originality* (or novelty) and *appropriateness*, i.e. relevance to the task at hand, and this two-criterion definition has become standard (e.g., Amabile, 1996; Feldman, Csikszentmihalyi & Gardner, 1994; Runco, 2004; Sternberg, 1988). The relative weighting of these two factors depends on the constraints and affordances of the task; in the arts, originality might be weighted more heavily, while for scientific or technological enterprises, appropriateness might be more important. *Surprise* is sometimes added as a third criterion (Boden, 2004). Some add *quality* as another criterion (Kaufman & Sternberg, 2007), while others use the term appropriateness in a way that encompasses quality.

*Process versus Product*

While to some it seems natural that creativity research focus on creative products, others view the creative process itself as more important than the product, stressing the therapeutic value of creativity (discussed below). In this view, the primary value of the creative process is that it enables the creator to express, transform, solidify, or unify the creator's understanding of and/or relationship to the world, while the external product provides a means of tracking or monitoring this internal transformation. This view is more prominent in eastern than western cultures.

Those who emphasize people and processes over products note that assessments of products involve social yardsticks that inevitably involve some degree of arbitrariness, subjectivity, and lack of knowledge. For example, defining creativity in terms of output originality and appropriateness implies that a particular work should lose its status as 'creative' if it is suddenly learnt that someone else came up with the same thing a decade earlier (it is no longer original), even though nothing has changed about the cognitive process that generated this work. It has been suggested that creative processes are those that *alter* how the individual conceives of or thinks about some aspect of their world, and that occur not merely due to assimilation of new sensory stimuli or knowledge but through *reflections on* those stimuli or learnings (Gabora, 2017). The external outputs are tangible pointers to such creative processes, and creative people are those who tend to engage in such processes and produce such outputs.

Others insist that any definition of creativity include such cognitive and personality characteristics as problem sensitivity, flexibility, the ability to analyze, synthesize, evaluate, and reorganize information, or cope with complexity. Creativity has even been defined as a complex or syndrome. In the end, there is probably no one-size-fits-all definition of creativity; nevertheless, these attempts to define it provide a departure point for further discussion and exploration.

### Personal versus Historical Creativity

Although the term 'creative' is often reserved for those who are famous for their creative output, daily life involves thinking things and doing things that, at least to some degree, have never been thought or done before, and as such, everyone is somewhat creative (Beghetto & Kaufman, 2007; Runco, 2004). Psychologists therefore distinguish between different kinds and degrees of creativity, such as between historical and personal creativity (Boden, 2004).

When the creative process results in a product that is new to humanity and makes an impact on the course of civilization, it is referred to as *historical creativity*. Historical creativity is also sometimes referred to as *eminent creativity*. When the creative process results in a product that is new to the creator, but someone else has come up with it before, or it is not creative enough to exert an impact on human civilization, it is referred to as *personal creativity*. Although personal creativity is not revolutionary, it can be a source of pleasure and amusement, as well as a gateway to more significant creative achievement.

A concept closely related to personal creativity is *everyday creativity* (Richards, 2004). Everyday creativity manifests in everyday life; it can come through in how one arranges furniture, prepares a meal, or interprets and shares experiences. Everyday creativity starts with taking an innovative, often unconventional approach to life that involves undertaking common tasks in uncommon ways, capitalizing on hidden opportunities, and finding unique solutions to challenges as they arise.

Historical and personal and creativity are sometimes called Big C creativity and Little C creativity, respectively. Some make the case for Mini C creativity, which involves making novel and personally meaningful interpretations of objects and events, and which may form the basis for more substantial creative acts (Beghetto & Kaufman, 2007).

### Distinguishing Creativity from Discovery and Invention

Creativity is related to, yet different from, discovery and invention. *Discovery* involves finding something already present and sharing it, e.g., Columbus' discovery of America. It could be said to be impersonal in the sense that it involves an aspect of life that everyone has potential access to, as opposed to deriving from personal experiences or feelings; thus, if one person hadn't discovered it, someone else would have (Moore, 1969). *Invention* involves unearthing something useful that did not exist before, e.g., the Wright brothers' invention of the airplane. Like discovery, it could be said to be impersonal in the sense that it does not emerge from the creator's unique personal experiences and feelings. Like discovery and invention, creativity involves unearthing and sharing something that was not present before. Some believe that for something to be creative, it must be profoundly personal, such that one feels the presence of a unique individual in the work. Thus, for example, when you look at hear Joni Mitchell's music, you get a sense of the person who created it.

### Creativity versus Intelligence

Early research suggested that creativity and intelligence are correlated up to an IQ of approximately 140, after which they diverge (Barron, 1963)**;** thus, someone can be very smart but not particularly creative, or highly creative but not particularly smart. Recent research suggests that the relationship between the two is more complex (Kim, 2008). Clearly it depends on how creativity and intelligence are being measured.

### Creativity versus Individual Learning

Creativity is central to cultural evolution, since it is what fuels cumulative, adaptive, open-ended cultural change. In the cultural evolution literature (e.g., Borenstein, Feldman, & Aoki, 2008), the creative processes resulting in cultural novelty are sometimes equated with 'individual learning,' but in fact they are not the same thing (Gabora, 2019). *Individual learning* deals with obtaining pre-existing information from the environment through non-social means (e.g., learning to differentiate different kinds of trees); the information already existed before the individual knew about it. In contrast, creativity involves the generation of new information, and it requires more than individual learning; it also involves *abstract thought*, i.e., the *reprocessing* of internally sourced mental contents. When this reprocessing results in the generation of useful or pleasing ideas, behavior, or artifacts that did not previously exist, it can be said to be creative.

## Historical Framework

In early times, a creative person was seen as an empty vessel that became filled with inspiration by a divine being. Psychologists initially paid little attention to creativity because it was thought to be too frivolous and complex for scientific investigation. Freud (1958) thought that creativity results from the tension between reality and unconscious wishes for love, sex, power, and so forth. While this view is less prominent now, Freud's notion of the *preconscious*—a state between conscious and unconscious reality in which thoughts are loose and vague yet interpretable—is still widely viewed as the source of creativity.

A pioneering effort toward demystifying creativity was Wallas' (1926) classification of the creative process into four stages. The first stage is *preparation,* which involves obtaining any relevant background knowledge about the creative task, and any instructions or past attempts or preconceptions regarding how to address it. This stage also involves conscious, focused work on the task. The second stage is *incubation*—unconscious processing of the task that continues while one is engaged in other tasks. The preparation and incubation stages may be interleaved, or incubation may not occur at all. Wallas proposed that following preparation and incubation, there is often a sudden moment of *illumination*, during which the creator has a new insight that may have to be worked and reworked in order to make sense. The idea at this point may be ill-defined or "half-baked", and require further reflection. Wallas' final phase is *verification*, which involves not just fine-tuning the work and making certain that it is correct, as the word implies, but putting it in a form that can be appreciated by others. The creative product may take the form of a physical object (e.g., a novel), or behavioral act (e.g., a comedy routine), or an idea, theory, or plan of action.

Today, creativity is of interest to many disciplines and approached from multiple directions. Even within the discipline of psychology, it is addressed in multiple ways. Cognitive psychologists study how people engage in cognitive processes considered creative, such as concept combination, problem solving, and analogy, and write computer programs that simulate these processes. Those who take a psychometric approach develop tests of creativity. Developmental psychologists examine creativity in children and throughout the lifespan. Social psychologists study how group dynamics, family dynamics, and cultural influences affect creativity. Clinical psychologists use art, dance, and music to improve wellbeing, self-understanding, and self-esteem. Neuroscientists investigate the biological basis of creativity. Organizational psychologists study creativity as it pertains to successful business strategies and entrepreneurship. Finally, cultural, comparative, and evolutionary psychologists address the question of how humans came to possess their superlative creative abilities, how these abilities compare to those of other species, how creativity differs across cultures, and in what sense creative ideas evolve over time.

## Creative Cognition

Cognitive approaches to creativity focus on the mental processes by which creative works are generated. These processes include concept combination, expansion of concepts, imagery, and metaphor. While most psychologists believe that creativity involves a combination of expertise, chance, and intuition, they differ in the degree to which they emphasize these factors. Expertise theorists point to evidence that Big C creativity is often achieved only through an extensive practise and learning. This is commonly referred to as the *10-year rule* on the basis of evidence that 10 years' experience in a domain is necessary for creative success (Ericsson, Krampe, & Tesch-Römer, 1993). Experts are more able than beginners to detect and remember domain-relevant patterns, and generate effective problem representations. Expertise theorists believe that creativity involves everyday thought processes such as remembering, planning, reasoning, and restructuring; they claim that no special or unconscious thought processes are required for creativity beyond familiarity with and skill in a given domain (Weisberg, 2006).

Others note that entrenchment in established perspectives and approaches may make experts more prone than beginners to phenomena such as functional fixedness, set, and confirmation bias. Some emphasize the role of chance meetings, opportunities or situations that invite creative ideation (Simonton, 1999). Still others view creativity as not so much a matter of expertise or chance but of listening to intuitions and following them through from an ill-defined state of potentiality to a well-defined state of actualization (Bowers, Farvolden, & Mermigis, 1995; Gabora, 2017). This view emphasizes how the association-based structure of memory provides a scientific basis for the phenomenon of intuition, and for findings that creative individuals tend to have flat associative hierarchies (Mednick, 1962), meaning they have better access to *remote associates:* items that are related to the subject of interest in indirect or unusual ways.

### *Stages and Processes in Creative Thinking*

Some of the earliest efforts to think systematically about creativity came from computer scientists, who viewed creativity as a process of heuristic search in which rules of thumb guide the inspection of different states within a state space (a set of possible solutions) until a satisfactory solution is found (Newell, Shaw & Simon, 1957). In *heuristic search,* the relevant variables of the problem or task are defined up front; thus, the state space is generally fixed. Examples of heuristics include breaking the problem into sub-problems, and working backward from the goal state to the initial state. It has since been proposed that creativity involves heuristics that guide the search for, not just a new possibility in a predefined state space, but a new state space itself (Boden, 1990; Kaplan & Simon, 1990).

Transformation of the state space is thought to be accomplished by switching from one representation of the task to another, sometimes referred to as *restructuring* (Weisberg, 1995). Restructuring is widely believed to be limited to the creative domain, sometimes referred to as the *problem domain*, e.g., for creative writing, the problem domain includes the space of possible plots, characters, and so forth. Others posit that restructuring is not limited to the problem domain in the sense that during creative thought none of the contents of mind is safeguarded from restructuring, arguing that this is why immersion in a creative task is potentially transformative (Gabora, 2017).

Restructuring is thought to involve generating new concept combinations, and viewing old concepts in new contexts (Ward, Smith, & Finke, 1999). Creative outcomes often require that restructuring be iterative and recursive, such that the output of one restructure becomes the input of the next (Basadur, 2016; Chan & Schunn, 2015). There is evidence that midway through creative processing, an idea may take different outward forms when reflected upon from different perspectives, and it consists of spontaneously merged (and potentially widely sourced) components (Scotney et al, in press). When the restructurings are not willy-nilly but reflect the structure of the network of understandings that constitute the creator's internal model of the world, or *worldview,* it is referred to as *honing* (Gabora, 2017). Honing may involve *self-organized criticality,* wherein small perturbations can have large effects; there is a chain reaction of insights as the implications of each new understanding is assimilated. The contents of the mind collectively self-organize, as evidenced by findings that it is possible for the domain-specific aspects of an idea to be stripped away such that it is amenable to re-expression in another form (Ranjan et al, 2014).

Wallas' classic four-stage model of creativity (discussed in Section [Historical Framework]) has been modified to include a phase that involves *problem finding* (noticing that something is amiss), problem posing (expressing the problem), problem construction (developing a detailed representation of the problem), or problem definition and redefinition (Amabile, 1996; Guilford, 1950). Indeed, Einstein famously claimed that real advance in science is marked by regarding an old problem from a new angle, and the formulation of a problem is often more challenging than its solution. Problem finding may involve sensing a gap or disturbing missing element (Torrance, 1963). Alternatively, it may take the form of a *seed incident* or *kernel idea* around which the creative project gradually takes shape. For example, studies of individual artists and designers have shown that their ideation process involves an ongoing interaction between artist and artwork during which the kernel idea goes from ill-defined to well-defined (Feinstein 2006; Locher 2010). Another modification of Wallas' model is the addition of a *frustration phase* prior to incubation, in which straightforward attempts to solve the problem are unsuccessful.

More often, the four stages of Wallas' original theory of creativity are reduced to two. The creative process has conventionally been conceived of as involving a divergent thinking phase followed by a convergent thinking phase. *Divergent thinking* was originally defined to be the kind of thought required for open-ended tasks, and assumed to involve the generation of multiple, often unconventional, ideas (Guilford, 1968; for a review see Runco, 2010). It was contrasted with *convergent thought,* which was originally defined as the kind of processing required for problems for which there is only one solution. More recently, divergent thought has been conceived of as reflecting on the task (or a prospective approach or solution to it) from unconventional contexts or perspectives, while convergent thought has been conceived of as reflecting on the task (or prospective approach or solution) from conventional contexts or perspectives.

This two-stage view is prevalent in scholarly theories of how the creative process works. In one well-known two stage theory, *Geneplore,* the generation of crudely formed *pre-inventive structures* is followed by exploration of them through elaboration and testing (Finke, Ward, & Smith, 1999). According to another two-stage theory, *Blind Variation Selective Retention* (formerly referred to as the Darwinian theory of creativity), new ideas come about through a trial-and-error process involving *blind* generation of ideational *variants* followed by *selective retention* of the most promising variants for development into a finished product (Simonton, 2013). The variants are said to be 'blind' in the sense that the creator has no subjective certainty about whether they are a step in the direction of the final creative product. Each variant is regarded as a separate entity, with its own probability and utility. BVSR implies that midway through the creative process the idea takes the form of multiple idea variants that exist simultaneously side by side as structurally distinct entities that can be individually assessed and selected from.

This classical conception of convergent and divergent thought assumes that ideas are like objects in the physical world that can be independently chosen and manipulated. Our everyday experience in a world of objects that exist in distinct locations and have definite boundaries may make it difficult to wean ourselves from the intuition that ideas in the mind behave this way as well. Not all scholars view creative thought as involving heuristic search or the generation of multiple discrete ideas followed by the selection and development of the most promising. Indeed, the term *search* implies that the idea was in there all along and one just has to find it, but memory does not work through a verbatim retrieval process; recalled items are unavoidably altered by what has been experienced since encoding, and spontaneously re-assembled in a way that relates to the task at hand (McClelland, 2011). One does not so much retrieve an item as reconstruct it (Brockmeier, 2010).

A more recent view (arising out of both research on the reconstructive nature of associative memory and mathematical models of how concepts combine and interact) is that creative thinking involves, not multiple ideas, so much as an idea that is 'raw' or ill-defined and therefore can be expressed many different ways and is subject to change (Gabora, 2017, 2019). In other words, due to its inherent ambiguity, it can manifest (i.e., actualize) as different kinds of outputs when viewed from different perspectives, like the different shadows cast by the same object when lit up from different directions. As the idea takes shape, this may *appear* to entail a shift to a new and different kind of mental processing, i.e., from 'generating many' possible mental representations of the unborn idea to 'refining one'. However, the real situation is much as if one went from shining light on an object from different directions—generating multiple shadows—to shining light on it from similar directions—making minor tweaks on the 'same' shadow. There is just one entity casting shadows, and just one process: the process of shining a light and casting shadows. Similarly, in creative cognition, there is just one mental representation of the unborn idea, and one process: gaining a clearer understanding of the idea and nudging it closer to its final form by looking at it in different ways. The issue of whether the mental representation of a creative idea is more like a physical object subject to search and selection or more like an unseen entity casting projections is not merely pedantic, for the formal structure of the two is entirely different (Gabora, 2005).

While the central aim of most theories of creativity is to account for the existence of creative products, the *honing theory of creativity* arose to account for the cumulative, open-ended nature of cultural evolution. It grew out of the view that humans possess two levels of complex, adaptive, self-organizing, evolving structure: an organismic level, and a psychological level (Barton, 1994). This psychological level can be referred to as a *worldview:* an individual's unique dynamic web of understanding that provides a way of both *seeing* the world and *being in* the world (i.e., a mind as it is experienced from the inside). In short, the worldview is the hub of a second evolutionary process—cultural evolution—that rides piggyback on the first—biological evolution—and that creative thinking fuels this second evolutionary process (Gabora, 2017). Creative acts and products render cognitive transformation culturally transmissible; thus, what evolves through culture is not creative contributions per se but worldviews, and creative outputs are vestiges of the worldviews that generated them.

The shifting between convergent and divergent thought is believed to involve different kinds of attention. *Defocused attention* involves diffuse activation of a broad region of memory such that new relationships can be perceived; this is conducive to the divergent thought processes in which new ideas are generated. *Focused attention* involves activating a narrow region of memory and treating items in memory as distinct chunks that can be readily operated on; it is conducive to the convergent thought processes in which creative ideas reach their final form.

### Domain Specific versus Domain General

Those who emphasize the role of expertise tend to view creativity as highly domain-specific; expertise in one domain is non-transferable to another (Baer, 2012). They note that expertise or eminence with respect to one creative endeavor is only rarely associated with expertise or eminence with respect to another; for example, creative mathematicians rarely become famous comedians. This is related to the above-mentioned evidence that 10 years' experience in a domain is necessary for creative success. Moreover, the existence of categorically distinct groups of creative achievement (as demonstrated by latent class analysis), furthers the argument for domain-specific creativity, especially in regard to creative products (Silvia, Kaufman, & Pretz, 2009).

*Domain-general theories* emphasize the generalizability of creative thinking across different domains, and the capacity for creators to connect different domains through associative thinking and metaphor (Hong & Milgram, 2010). Research that focuses, not on the *products* resulting from creative thought, but on the *inspiration phase* of the creative process suggest that creativity is more domain-general than is widely believed (Scotney et al, 2019). For example, cross-domain influences such as artistic avocations can stimulate creativity in scientists (Root-Bernstein, 2003), and that when people express themselves in different creative domains these outputs bear a recognizable style or 'voice' (Ranjan et al, 2014). In another study, when painters were instructed to paint what a particular piece of music would 'look like' if it were a painting, naïve participants were able to correctly identify at significantly above chance which piece of music inspired which painting. Although the medium of expression was different, something of its essence remained sufficiently intact for people to detect a resemblance between the new creative output and its inspirational source. Such findings suggest that the creative mind seeks to explore and express its distinctive structure and dynamics using whatever means available.

The view that creativity is domain-general is also supported by studies involving creativity checklists, self-report scales, and other sorts of psychometric or personality data (Plucker, 1998). The relevance of these studies to the general versus specific debate has been questioned because they do not measure creative outputs, but traits associated with creativity. However, those who stress process over product claim that these data tell us about the internal, less tangible but equally important counterpart to the external creative outputs. If we ask not, 'are individuals talented in multiple creative domains?' but, 'can individuals use multiple creative domains to meaningfully develop, explore, and express themselves?' the answer is likely to be affirmative. In short, little-c creativity in multiple domains, and even if big-C creators work primarily in a single domain, they are often employing cross-domain thinking when they create.

Many scholars currently espouse a less dichotomous view of creativity that incorporates both domain-specific and domain-general aspects (Gabora, 2017; Plucker & Beghetto, 2004). Possible mechanisms underlying cross-domain creativity have been suggested (Simonton, 2009;

Sternberg, 2009), and tested in empirical studies (Boccia et al., 2015; An & Runco, 2016; Barbot, Besancon, Lubart, 2016; Palmiero et al., 2019). In short, creativity in one domain may help but not guarantee creativity in another.

## Individual Differences, Case Studies, and Clinical Approaches

In the early days of psychology, experiments in strictly controlled laboratory conditions were encouraged; case studies of individuals, and particularly introspective accounts, were not taken seriously. Recently, researchers have acknowledged the artificiality of many laboratory studies. There is stronger appreciation for taking an ecological approach, which involves studying people in their everyday environments engaged in everyday tasks, and treating individual differences not just noise, but as interesting in their own right. Individual differences are particularly important in the study of creativity. Thus, although case studies and introspective accounts are flawed and can never take the place of more controlled approaches, they have a place in the effort to achieve a multifaceted understanding of how the creative process works.

Clinical studies, and historiometric studies, which involve statistical analyses of data that comes from analysis of the biographies, case studies, and so forth, of large numbers of big-C creators, provide mixed evidence that creative individuals are more emotionally unstable and prone to affective disorders such as depression and bipolar disorder (Taylor, 2017). Clinical approaches to creativity are not exclusively focused on the negative; they also investigate how creative therapies such as art therapy, music therapy, and dance therapy can enhance wellbeing and coping skills and strengthen personal identity (Knill, Levine, & Levine, 2005).

## Developmental Antecedents and Personality Attributes of Creative People

Developmental approaches to creativity focus on creativity in children and throughout the lifespan. An important set of findings to come from this research concerns the relationship between emotions and creative play during childhood. The extent to which children access affect-laden (emotional) thought is correlated with ratings of cognitive flexibility, associative thinking, and creativity, and the extent to which children engage in fantasy and play (Russ, 1993). In addition, the degree of fantasy and imagination at ages six to seven was related to the divergent thinking ability in high school.

Personality approaches to creativity focus on the psychological traits associated with creativity. There does indeed appear to be something to the notion of a creative personality type (Batey & Furnham, 2006; Eysenck, 1993; Feist, 1998; Martindale & Daily, 1996), Creativity is correlated with self-confidence, aesthetic orientation, independence of judgment, risk-taking, openness to experience, tolerance of ambiguity, impulsivity, lack of conscientiousness, high energy, attraction to and ability to handle complexity, problem sensitivity, flexibility, the ability to analyze, synthesize, evaluate, and reorganize information, engage in divergent thinking. Creative individuals differ with respect to whether they are internally versus externally oriented, person-oriented or task-oriented, and explorers (who tend to come up with ideas) or developers (who excel at turning vague or incomplete ideas into finished products).

On the basis of the above-mentioned findings that creative individuals are more prone to anxiety and emotional (affective) disorders as well as suicide and substance abuse, creativity is sometimes said to have a dark side (Cropley, Cropley, Kaufman, & Runco, 2010). However, since we all benefit from the creativity of a few by imitating, admiring, or making use of their creative outputs, it is not necessary for everyone to be creative. Indeed, excessive creativity may result in reinventing the wheel, and absorption in ones' own creative ideas may interfere with assimilation or diffusion of proven effective ideas. Computer modeling suggests that society self-organizes to achieve a balance between relatively creative and uncreative individuals (Gabora & Tseng, 2017). The social discrimination that creative individuals often endure until they have proven themselves may aid in achieving this equilibrium.

## Measuring Creativity

The *psychometric approach* to creativity involves developing creativity tests and methods for improving creativity. The most widely known creativity test the *Torrance Test of Creative Thinking* (Torrance, 1962, 1996). It consists of multiple components or sub-tests. Perhaps the most well-known of these is the *Unusual Uses Test* in which participants are asked to think of as many uses for a common object (e.g., a brick) as possible. Another is the *Product Improvement Test*, in which participants are asked to list as many ways as they can to change a product to make it more useful or desirable, (e.g., to change a toy monkey so children will have more fun playing with it).

Although these tests are still in use, they are criticized because they do not test adequately for the ability to engage in the kind of iterative honing process required for big-C creativity. Another test that is superior in this regard is Amabile's (1996) Consensual Assessment Technique, in which a panel of expert judges is asked to rate the creativity of products in their area of expertise, e.g., collage making, story writing, or poetry. Still other methods involve step-by-step case studies of actual creators at work (e.g., Choi & DiPaola, 2013; Feinstein 2006; Locher 2010).

The most well-known technique for *improving* creativity is *brainstorming,* which takes place in groups, and involves encouraging them to suggest as many ideas as possible, no matter how seemingly 'crazy', while discouraging criticism of ideas (Rawlinson, 2017). These techniques are used by businesses and organizations to develop successful business strategies, and in business schools to foster entrepreneurship.

## The Impact of Environment

Creativity is affected by environmental conditions. Certain individual situations, such as education and training, role models and mentors, and perhaps surprisingly, childhood trauma, are correlated with historical creativity (Simonton, 1999). Social and cultural approaches examine how family dynamics, group dynamics, and cultural influences, affect creativity, as well as how creativity compares across different cultures. Maslow (1974) believed that creativity is fostered by environments that are supportive and free of evaluation, which he claimed are conducive to self-actualization. However, positive social environments do not necessarily cultivate creativity. For example, highly creative people tend to experience a lack of parental warmth, and are more likely to have experienced the death of a parent at an early age, and raise fewer children, than average. According to the systems approach, creativity occurs through an interaction between (1) the individual, i.e., the creator, (2) a field which is set of relevant individuals in society, i.e., the people involved in same creative domain as the individual, and (3) the domain, i.e., the set of relevant ideas (Csikszentmihalyi 1997; Sawyer 2006).

## Neural Basis for Creativity and Brain Regions Involved

Biological approaches investigate the underlying neural and physiological mechanisms underlying creativity, and the extent to which there is a genetic basis to creativity. Twin studies and other sorts of evidence suggest creative abilities are, at least to some extent, genetically inherited (Eysenck 1995). One way to investigating the brain mechanisms underlying creativity involves dissecting the brains of people who were particularly creative during their lifetimes. It has been shown that Einstein's brain had (1) a partially absent Sylvian fissure, which may have facilitated communication between different parts of the brain, and (2) a high ratio of glial cells to neurons in both area 9 of prefrontal cortex, which is associated with planning, attention, and memory, and area 39 of the left inferior parietal cortex, which is associated with synthesizing information from other brain regions.

A less dramatic but more common way to investigate the neural underpinnings of creativity involves examining brain activity when people engage in creative activities using electroencephalography (EEG) or functional magnetic resonance imaging (fMRI). Though application of such methods is stymied by the fact that many brain areas are active during creative thought, much has already been learned (Beaty, Seli, & Schacter, 2019; Vartanian & Jung, 2018). There is evidence that different kinds of creativity (deliberate versus spontaneous, and emotional versus cognitive) involve different neural circuits. Creative thought appears to be facilitated by lower levels of noradrenaline and dopamine—catecholamines directly linked to cognitive control, prefrontal functioning, and cortical arousal. EEG experiments show that divergent thinking tasks produce decreased beta range synchrony and increased alpha range synchrony over the frontal cortex, which suggests there is a loosening of executive control and lower prefrontal cortical arousal during creative thought. There is also indirect neuroscientific support for the notion that creativity involves the ability to tune one's mode of thought to the task. Prior to finding the solution to an insight problem there is neural recruitment of the prefrontal and executive memory networks, as well as of the so-called 'default network' that is active when one engages in spontaneous mind wandering, which suggests that mind wandering can have a utilitarian function (Fox, & Christoff, 2018).

Physiological research into creativity has revealed evidence of an association between creativity and high variability in physiological measures of arousal such as heart rate, spontaneous galvanic skin response, cortical activity, and EEG alpha amplitude (Jausovec & Bakracevic, 1995). For example, although creative people tend to have higher resting arousal levels, when engaged in creative problem solving they tend to have lower than average arousal levels These findings in conjunction with the cognitive findings discussed previously suggest that during creative activities, creative individuals are particularly prone enter a state that is quite different from their normal resting state, a state that has both a physiological aspect (low arousal level) and a cognitive aspect (divergent or associative mode of thought).

We have also learned something about what is going on during creative thought at the level of neurons. The fact that memory distributed (encoded items are spread out across multiple neurons) and content-addressable (there is a systematic relationship between the content of an item and the location of the neurons involved in its encoding) means that knowledge and ideas that are relevant to the task at hand to come to mind naturally without systematic search (Gabora, 2017). An unfinished or 'half-baked' idea may be what results when remotely related items encoded in narrowly overlapping distributions of neural cell assemblies get evoked simultaneously. The vagueness of such an idea, and the sense that it holds potential, as well as its capacity to actualize in different ways depending on how one thinks it through, may be side effects of interference. *Interference* refers to the situation wherein a recent memory interferes with the capacity to recall an older memory because they are encoded in overlapping distributions of neurons in the brain. It is generally thought of as detrimental, but it may be beneficial with respect to creativity. When two or more items encoded in overlapping distributions of neural cell assemblies interfere with each other and get evoked simultaneously, a new idea may be the result. The phenomenon of interference leading to creative ideation can be referred to as *creative interference*. The vagueness of the new idea may reflect that it is uncertain how, in the context of each other, the interfering components come together as a realizable whole. The ability to work with an idea in this state is related to the personality trait of tolerance of ambiguity. The above-mentioned evidence that creativity is at least partly domain-general is consistent with high levels of functional connectivity amongst the default, salience, and executive neural circuits in creative individuals during creative thinking (Beaty et al., 2018).

**The Evolution of Human Creativity**

The human lineage was not always so creative as we are now (Asma, 2017). Comparative and evolutionary approaches address how humans evolved their superlative creative abilities, how these abilities compare with those of other species, and in what sense ideas can be said to evolve.

The earliest preserved signs of our creativity include primitive stone tools over three million years ago (Harmand et al., 2015). Since this corresponds with an increase in brain size, it has been suggested that this enabled memories to be encoded in more detail, such that there were more ways in which one experience could evoke a reminding of another. This provided more ways of chaining and recombining thoughts and experiences into an integrated understanding of the world, which both enabled and constrained the generation of creative ideas. Computational modeling has provided support for this hypothesis (Gabora & Smith, 2018).

The Middle Upper Paleolithic has been referred to as the "big bang of creativity" for it marks the beginnings of art, science, and religion (Mithen, 1998). It has been suggested that this is due to onset of the capacity to spontaneously shift between divergent and convergent modes of thought by shrinking or expanding the field of attention. Once it was possible to tailor one's mode of thought to the demands of the situation, the fruits of one mode of thought could be used as ingredients for the other, resulting in a richer understanding of the world and enhanced potential to creatively change it. Tasks that required either mode of thought, or both at different stages of the creative process, could be carried out more effectively. The last few hundred years have witnessed another accelerating surge of creativity, particularly in the realm of technology, but with environmental degradation and resource depletion as unfortunate consequences.

Another evolutionary approach uses societies of artificial agents that invent ideas and imitate neighbors' ideas, to understand how ideas evolve over time. This approach has provided evidence that the evolution of ideas gives rise to some phenomena observed in biological evolution, such as (1) an increase in complexity and fitness (usefulness) over time, and (2) an initial increase in diversity as the space of possibilities is explored followed by a decrease associated with convergence on the fittest possibilities. If even a small fraction of agents are creative, new ideas spread in waves by imitation throughout the society, reaching other creators who puts another spin on them. Thus, over time they accumulate change and evolve. The more creative the creative agents are, the fewer of them there must be, so as to maximize the evolution of fitter ideas, suggesting that more creativity is not always better; as in biological evolution, creativity must be balanced with continuity (Gabora & Tseng, 2017). In collaboration with archaeologists, creativity researchers also trace the process by which one technological innovation paved the way for another (Dasguptas, 1996). Thus, evolutionary theories and computational models can assist the process of organizing human-created artifacts into cultural lineages.

## Computational Creativity

We saw that computational modeling is particularly useful to those who investigating hypotheses concerning the mechanisms by which humans became creative. Because the onset of such mechanisms left no detectable trace, one begins by establishing which hypotheses are at least computationally feasible. However, researchers studying the evolution of creativity are not the only ones to take a computational approach; indeed, efforts to develop computational models of creativity date back to the 1950s and '60s. Herbert Simon developed a computer program called BACON that came up with scientific laws. The nineties and early 2000s have witnessed a plethora of computer programs that make use of sophisticated machine learning techniques such as deep learning and evolutionary computation to generate music (Agres, Forth, & Wiggins, 2016), art (DiPaola & Gabora, 2009; DiPaola, & Gabora, & McCaig, 2018), and linguistic forms such as stories and humor (Veale, 2016), some of which can generate results that feel 'human-like' (Mekern, Hommel, & Sjoerds, 2019). Computational models have also been used to model the shifting between two distinct modes of thought in the creative process (Martindale 1995; Gabora & Smith, 2018; Helie & Sun, 2010), as well as incubation and insight (Thagard & Stewart, 2011).

## Implications for Policy and Practice

The psychological study of creativity has implications for theory, policy, and practice in a number of areas. A first area of application is clinical. Creative activities such as art making, music making, dance, and drama are increasingly seen to have therapeutic effects that can be effective in both clinical and non-clinical settings. The transformation that occurs on canvas or on the written page is often mirrored by a potentially therapeutic sense of personal transformation and self-discovery from within. Immersion in the creative task has been referred to as a state of flow that may share characteristics with spiritual or religious experiences.

A second, related area of application is education and childrearing. Both teaching to *foster* creativity, such as by asking open-ended questions, and *creative teaching methods*, such as learning about food webs by painting murals, may help prepare students to thrive in an everchanging and increasingly uncertain world (Snyder, Gregerson, & Kaufman, 2012). Free time, in which children can develop imagination and a sense of personal identity as they explore where their own free-association processes lead them is also important; creative play in childhood facilitates access to affect-laden (emotional) thoughts, which may enhance cognitive flexibility and divergent thinking abilities (Russ, 2013). Research on *intrinsic motivation* showed that rewards for creative work may in fact inhibit creativity because focusing on an external reward leads people to neglect the internally rewarding nature of creative acts (Amabile, 1996).

A third area of application is in business settings. For example, psychological work on brainstorming (mentioned above), in which people get together as a group and put forward ideas in an open and accepting environment, has shown that it may be more effective when group work is followed immediately by individual work, or when individuals communicate by writing so as to avoid the problem of everyone talking at once.

## Conclusion and Future Directions

Our creativity distinguishes humans from other species and it has radically transformed the planet we live on. The study of creativity is an exciting area that brings together many different branches of psychology: cognitive, social, personality, developmental, organizational, clinical, neuroscience, mathematical models, and computer simulations. Past and current areas of controversy concern the relative contributions of expertise, chance, and intuition, whether the emphasis should be on process versus product, whether creativity is domain-specific versus domain-general, and the extent to which creativity is correlated with affective disorders. Promising areas for further psychological study of creativity include computational modeling, and work on the biological basis of creativity, and studies that track specific creative ideation processes over time. It is an important time to study creativity because with the accelerating pace of cultural and environmental change we are ever more in need of new solutions and perspectives, and new possibilities for exploring and expressing our ever-changing identities.

## Acknowledgements

This work was supported by a grant (62R06523) to the first author from the Natural Sciences and Engineering Research Council of Canada.